\documentclass[aip,apl,amsmath,amssymb, numerical,reprint,twocolumn]{revtex4-1}
\usepackage{hyperref}
\usepackage{siunitx}
\DeclareSIUnit\sq{\ensuremath{\Box}}
\usepackage{chemformula}
\usepackage{xcolor,colortbl}

\usepackage{graphicx}% Include figure files
\usepackage{dcolumn}% Align table columns on decimal point
\usepackage{bm}% bold math
\usepackage[utf8]{inputenc}
\usepackage[T1]{fontenc}
\usepackage{mathptmx}
\usepackage{etoolbox}

\makeatletter
\def\@email#1#2{%
 \endgroup
 \patchcmd{\titleblock@produce}
  {\frontmatter@RRAPformat}
  {\frontmatter@RRAPformat{\produce@RRAP{#1\href{mailto:#2}{#2}}}\frontmatter@RRAPformat}
  {}{}
}%
\makeatother

\begin{document}

%\preprint{}

\title{Fabrication and characterization of vacuum-gap microstrip resonators} 

\author{Christian Schlager}
\email[]{christian.schlager@uibk.ac.at}
\affiliation{University of Innsbruck, Institute for Experimental Physics, Innsbruck, Austria}
\affiliation{Institute for Quantum Optics and Quantum Information, \\ Austrian Academy of Sciences, Innsbruck, Austria}

\author{Romain Albert \thanks{Current Address: Silent Waves, 69-73 rue Félix Esclangon, 38000 Grenoble, France}}
\affiliation{University of Innsbruck, Institute for Experimental Physics, Innsbruck, Austria}
\affiliation{Institute for Quantum Optics and Quantum Information, \\ Austrian Academy of Sciences, Innsbruck, Austria}
\affiliation{Current Address: Silent Waves, 69-73 rue Félix Esclangon, 38000 Grenoble, France}

\author{Gerhard Kirchmair}
\affiliation{University of Innsbruck, Institute for Experimental Physics, Innsbruck, Austria}
\affiliation{Institute for Quantum Optics and Quantum Information, \\ Austrian Academy of Sciences, Innsbruck, Austria}

\begin{abstract}
In traveling-wave parametric amplifiers (TWPAs), low-loss capacitors are necessary to provide \SI{50}{\ohm} impedance matching to the increased inductance that is brought in by the nonlinear elements used for amplification, be it Josephson junctions or high kinetic inductance materials. Here we report on the development of a fabrication process for vacuum-gap microstrips, a design in which the ground plane is suspended in close proximity above the center conductor without the support of a dielectric. In addition to high-capacitance transmission lines, this architecture also enables air-bridges and compact parallel-plate capacitors. The performance of the fabrication is examined using distributed aluminum and granular aluminum resonators in a cryogenic dilution refrigerator setup, showing quality factors on par with the fabrication processes used in state-of-the-art TWPAs. In addition to characterizing the dependence of the quality factors on power, also their behavior with respect to temperature is explored, applying a model based on thermal quasi-particles and saturable two-level systems (TLS), showing that the quality factors of the resonators are limited by TLS.

\end{abstract}


\maketitle %\maketitle must follow title, authors, abstract and \pacs

    Implementing low-footprint capacitance while simultaneously keeping dielectric losses low is a challenge in fabricating superconducting resonators\cite{zmuidzinas_superconducting_2012} and circuits in general. Especially traveling-wave parametric amplifiers (TWPAs) using Josephson junctions or high kinetic inductance materials have a need for large, compact, and low-loss capacitors to achieve impedance matching, which is necessary for preventing unwanted reflections\cite{esposito_perspective_2021}. 
    
    While the first implementations of TWPAs\cite{ho_eom_wideband_2012,bockstiegel_development_2014} made use of impedance transformers on both ends of the amplifier, state-of-the-art designs either use interdigitated coplanar capacitors \cite{adamyan_superconducting_2016,malnou_three-wave_2021} or parallel-plate like geometries with a dielectric layer in between \cite{macklin2015,shan_parametric_2016,klimovich_demonstration_2023,ranadive_traveling_2024,ranadive_kerr_2022,planat_photonic-crystal_2020,fadavi_roudsari_three-wave_2023} to achieve \SI{50}{\ohm} matching. Coplanar designs are usually lower in loss, due to the lower participation of surfaces, interfaces, and non-crystalline dielectrics. However, achieving large capacitance in this geometry requires large-footprint structures and dedicated addressing of slot-line modes between the individual separated parts of the ground plane. Parallel-plate capacitors offer large capacitance with a compact footprint and an uninterrupted ground plane. However, TWPA devices using this design are often suspected to be limited in their noise performance by the dielectric loss introduced by the amorphous dielectrics in the capacitors. Realizing a microstrip with a vacuum-gap instead of the conventional lossy dielectric layer between the two conductors is therefore a promising approach for realizing compact capacitors while maintaining comparably low losses.
    
    Vacuum-gap air-bridges are frequently used to bridge over planar structures and ensure ground plane connectivity over a chip divided by coplanar waveguides\cite{abuwasib_fabrication_2013,chen_fabrication_2014,janzen_aluminum_2022,jin_fabrication_2018}. However, they are usually designed with heights of micrometers to not significantly alter the impedance of the transmission line they are spanning. To significantly increase the capacitance of the transmission line, the vacuum-gap structure has to be suspended at distances of hundreds of nanometers or less. This however, comes at the cost of more difficult fabrication as the narrower gap is harder to access and increases the risk of collapse. Narrow vacuum gaps have been achieved using sacrificial layers of \ch{Si}, \ch{Nb} or \ch{SiN_x} \cite{cicak_vacuum-gap_2009,cicak_low-loss_2010} or \ch{Al} \cite{boussaha_development_2020}. For removing \ch{Si}, \ch{Nb} and \ch{SiN_x} \ch{SF6} plasma etching, and for \ch{Al} a bath in a basic solution is used. The former method is incompatible with \ch{Nb}, \ch{NbN}, and \ch{NbTiN}, while the latter is incompatible with \ch{Al}, granular aluminum (grAl), and Josephson junctions.
        
        \begin{figure*}
            \centering
            \includegraphics[width=1.\textwidth]{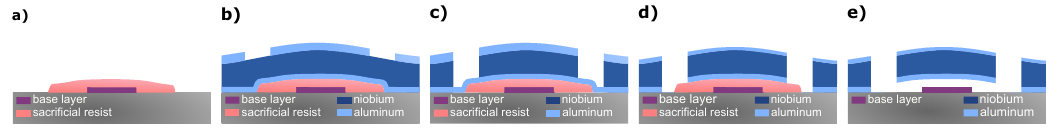} 
            \caption{Schematic depiction of the fabrication process. \textbf{(a)} An \SI{80}{\nano\metre} thick layer of resist covers the base layer, protecting it and acting as a sacrificial placeholder. \textbf{(b)} The triple-layer ground plane is deposited on top with holes patterned in the topmost aluminum layer. \textbf{(c)} The aluminum mask enables the etching of access holes through the niobium layer using a \ch{CF4} + \ch{O2} plasma RIE. \textbf{(d)} Subsequently we remove the innermost aluminum layer using a \ch{BCl3} + \ch{Ar} plasma. The plasma also attacks the outermost aluminum layer, but does not remove it entirely. \textbf{(e)} Through the holes, the sacrificial resist is removed by organic solvents before the vacuum-gap microstrip is released by critical-point drying. Dimensions in horizontal and vertical direction are not to scale. Typical widths of the sacrificial resist used for the samples in this work are around \SI{10}{\micro\metre}, while the thickness of the sacrificial resist is \SI{80}{\nano\metre}.}\label{fig:fabsketch}
        \end{figure*}
  
    The fabrication process described in this work uses a thin film of resist as a sacrificial layer that is afterwards removed by organic solvents and subsequent critical-point drying to create a vacuum-gap (VG) microstrip. This increases the compatibility to a wide range of different materials within the same design, including the materials mentioned above. The nominal gap of the VG microstrip is \SI{80}{\nano\meter}. While comparable sizes have been achieved with sacrificial layers made from \ch{Si}\cite{jong_measurement_2025}, our process produces the smallest vacuum gap achieved with a resist as the sacrificial layer. We have fabricated aluminum and granular aluminum resonators coupled to the same transmission line, assessed their quality factors, and also explored their behavior with temperature.

        \begin{figure}
            \centering
            \includegraphics[width=1.\linewidth]{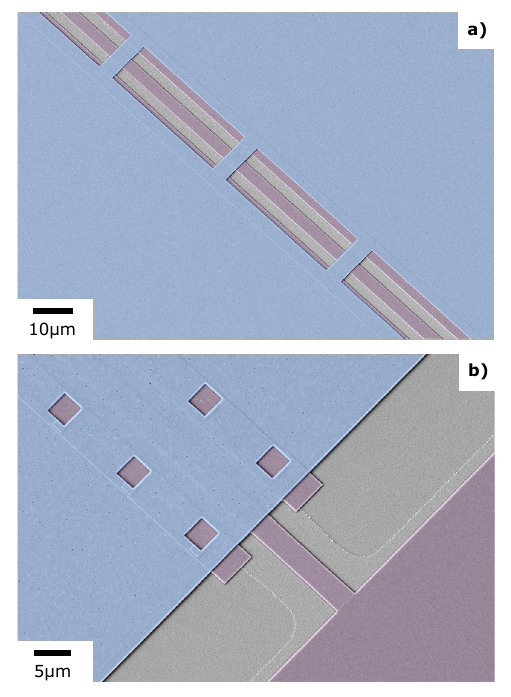} 
            \caption{SEM false color images of different vacuum-gap microstrip structures. Analogous to figure \ref{fig:fabsketch}, purple indicates the base layer, blue the ground plane, and gray the substrate. \textbf{(a)} VG air-bridges over a CPW transmission feedline. \textbf{(b)} VG microstrip with access holes through the ground plane starting from a bonding pad (lower right corner). The two strips parallel to the center conductor enhance the adhesion of the sacrificial resist to the substrate. The dark spots that can be seen on the ground plane are due to superficial damage done to the outermost aluminum layer by the last etching step. In both images, the sacrificial resist has been removed; the outline of where it was can be seen in \textbf{(b)} where part of the silicon surface was protected by the resist during the last etching step. }
            \label{fig:Imaging} 
        \end{figure}

        In Fig.\,\ref{fig:fabsketch} we sketch the full fabrication process. The first step is the definition of the base layer on a silicon substrate (see Fig\,\ref{fig:fabsketch}\,(a)). In the resonator samples fabricated, the base layer contains sputtered granular aluminum\cite{rotzinger_aluminium-oxide_2017}(\SI{20}{\nano\metre}) and evaporated aluminum (\SI{40}{\nano\metre}), patterned and deposited in subsequent steps. First, the granular aluminum is deposited by sputtering in an \ch{Ar} + \ch{O2} atmosphere (see Supplementary Material \ref{sec:appendix_grAlSputtering}). We choose a kinetic inductance of approximately \SI{100}{\pico\henry\per\sq}. Then, aluminum is deposited in a separate lift-off process. As a spaceholder for the future vacuum-gap, the wafer is coated with an \SI{80}{\nano\meter} thick sacrificial layer of the negative resist ma-N 2401, which is patterned using electron-beam lithography (see Fig.\,\ref{fig:fabsketch}\,(a)). The resist is overdosed to obtain rounded edges. To guarantee sufficient adhesion of the sacrificial resist as well as a well-defined gap between center conductor and ground-plane, additional metal strips in the base layer are used (visible in Fig.\,\ref{fig:Imaging}\,(b)). The metasilicate-based developer ma-D 377 is used instead of TMAH-based alternatives to ensure compatibility with aluminum.

        The patterned sacrificial resist is covered by a double-layer ground plane consisting of an inner layer of evaporated aluminum (\SI{10}{\nano\metre} thick) and the sputtered niobium main body (\SI{300}{\nano\metre} thick). The sputtering parameters are calibrated to produce a strain-free film of \ch{Nb} which is necessary for the structural integrity of the VG-microstrips. The reason behind choosing the described combination of niobium and aluminum as the ground plane is twofold: Firstly, the strain in the sputtered film of niobium can be controlled via the sputtering parameters. Secondly, aluminum and niobium can be selectively etched, which is relevant for the etching process described in the next paragraph. The depositions take place in a combined evaporation and sputtering apparatus without breaking vacuum in between. To ensure electrical contact to the base layer, the aluminum evaporation is preceded by \ch{Ar} ion milling to remove any native oxides. Finally, a patterned layer of aluminum is deposited through a lift-off process (see Fig.\,\ref{fig:fabsketch}\,(b)).
    
       The niobium layer of the ground plane is etched using \ch{CF4} + \ch{O2} reactive ion etching (RIE). This chemistry is highly selective between the aluminum mask and the niobium main layer. The etching stops at the innermost thin layer of aluminum, preventing any risk of etching into the sacrificial resist and potentially harming the base layer (see Fig.\,\ref{fig:fabsketch}\,(c)). Finally, the inner layer of the ground plane is etched in the same RIE using a combined \ch{BCl3} and \ch{Ar} plasma (see Fig.\,\ref{fig:fabsketch}\,(d)). The plasma causes some damage to the topmost aluminum layer, but it is not sufficient to etch it completely.
    
        We deploy a sequence of organic solvents (N-Methyl-2-pyrrolidone (NMP), acetone, and ethanol) to remove the sacrificial resist and release the ground plane (see Fig.\,\ref{fig:fabsketch}\,(d)). To ensure optimal removal of the sacrificial resist, the sample is left for two days in NMP, one day in acetone, and one day in ethanol. Without letting the wafer dry at any point during this process, we transfer it in ethanol to a critical point dryer to release the ground plane, similar to MEMS fabrication processes\cite{jafri_critical_1999}.  This step is essential for creating such narrow vacuum gaps with this technique, as surface tension during drying would otherwise cause the collapse of the structures. In Fig.\,\ref{fig:Imaging}, two SEM images augmented with false colors illustrate air-bridges over a CPW transmission line and the start of a VG microstrip, both implemented using the described process.

        \begin{figure*}
            \centering
            \includegraphics[width=\textwidth]{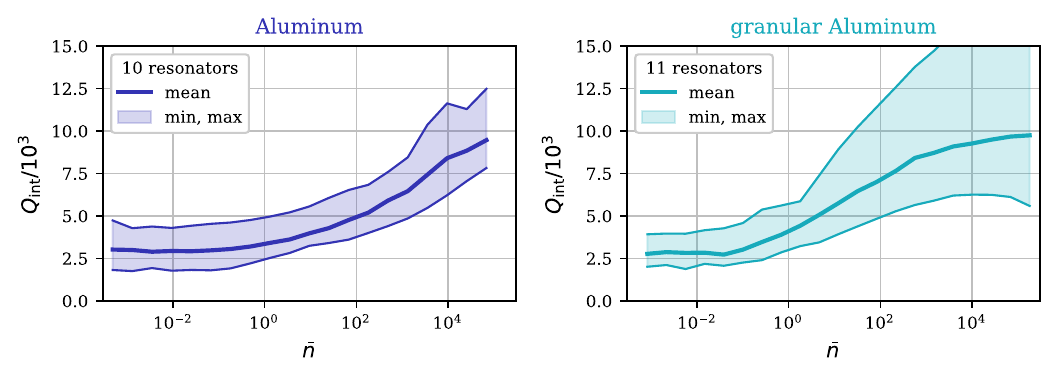} 
            \caption{Internal quality factors of aluminum \textbf{(left)} and granular aluminum \textbf{(right)} vacuum-gap microstrip resonators over average photon number $\Bar{n}$. The quality factors were obtained by fitting the resonant responses in the VNA transmission data of the feedline. The data of all resonators is contained within the shaded regions, while the line in the center represents the average. For the raw data of the individual resonators, see Supplementary Material \ref{sec:appendix_QualityFactors}.}
            \label{fig:Q_over_n}
        \end{figure*}
        
        To examine the microwave performance of the presented fabrication process, $\lambda/4$ resonators using the VG-microstrip structure were fabricated in a hanger configuration along a transmission line. The transmission line is implemented as a coplanar waveguide with air-bridges every \SI{50}{\micro\metre} for good electrical connection of the ground plane (see Fig.\,\ref{fig:Imaging}\,(a)). The specific interval is chosen to be smaller than the wavelength of the microwave signals to avoid band gaps introduced by the periodic capacitive loading. Multiple $\lambda/4$ resonators in full (close to \SI{100}{\percent} coverage) vacuum-gap microstrip architecture (see Fig.\,\ref{fig:Imaging}\,(b)) are coupled to this transmission line. Their center conductors are made of two different materials, aluminum and granular aluminum. The coupling is done with parallel-plate vacuum-gap capacitive couplers. The samples are wire-bonded to a PCB in a copper sample holder and placed inside a dilution refrigerator with a base temperature of \SI{20}{\milli\kelvin} (see Supplementary Material \ref{sec:appendix_Setup}). A vector network analyzer (VNA) is used to measure the transmission of the feedline.
        
        Two samples were cooled down, with Chip 1 containing two and Chip 2 containing one feedline. Each feedline contains 5 VG microstrip resonators with an aluminum center conductor and 4 VG microstrips with a center conductor made from grAl. Of the total of 15 aluminum and 12 grAl resonators, 10 aluminum and 11 grAl resonators could be observed in the transmission spectrum (see Supplementary Material \ref{sec:appendix_QualityFactors}). The quality factors of all 21 resonators were extracted by performing resonance circle fits \cite{probst_efficient_2015,gao_physics_2008,deng_analysis_2013}. This method allows us to extract the coupling and internal quality factor of a resonator by fitting a circle to the complex response of the system when probed near resonance. Fig.\,\ref{fig:Q_over_n} shows the extracted internal quality factors $Q_\text{int}$ of the aluminum and granular aluminum resonators. The x-axis shows the average number of photons $\Bar{n}$, populating the resonator on resonance. It is calculated from the internal and coupling quality factors and the incident power $P_\text{inc}$ in \SI{}{\watt} at the sample via \cite{bruno_reducing_2015}
            \begin{equation}
                \Bar{n} = \frac{2Q_\text{l}^2}{Q_\text{c}\hbar\omega_\text{res}^2} P_\text{inc},
            \end{equation}
        where $Q_{l}$ denotes the loaded and $Q_{c}$ the coupling quality factor of the resonator, $\omega_\text{res}$ is the angular resonance frequency. The incident power at the sample is obtained by estimating the total attenuation of the input chain (see Supplementary Material \ref{sec:appendix_Setup}).
        
        The quality factors for low powers can be seen towards the left part of the diagrams in Fig.\,\ref{fig:Q_over_n}, where the curves flatten out. With increasing power, i.e.\ higher average photon number in the resonator, the internal quality factors increase due to saturation of losses connected to two-level systems (TLS).
        
        Considering the application of our VG microstrips in TWPAs, we are especially interested in the low-power part of the losses, as they are essential for the noise added to a small signal propagating through the amplifier. The losses for high powers are important for the attenuation suffered by a strong pump tone or the amplified signal propagating through the system, thus limiting the achievable gain. From the low-power section of Fig.\,\ref{fig:Q_over_n} we read quality factors of $Q_\text{int} \approx \SI{3000}{}$, or loss tangents of $\tan\delta = Q_\text{int}^{-1}\approx\SI{3e-4}{}$ respectively. 
        
        It has to be noted, that previous works on vacuum-gaps, using \ch{Si}, \ch{Nb} or \ch{SiN_x} \cite{cicak_vacuum-gap_2009,cicak_low-loss_2010} or \ch{Al} \cite{boussaha_development_2020} as a sacrificial layer, have seen higher quality factors for their capacitors of $Q_\text{int}=\SIrange[]{2.5}{3.3e4}{}$ or $Q_\text{int}=\SI{4.3e5}{}$ respectively. We attribute the limiting loss of our devices to resist residue that remains on the surfaces inside the VG-microstrips (see Supplementary Material \ref{sec:appendix_EDX}). It is worth noting that the vacuum-gap in this work is by a factor of \SIrange{2}{4}{} smaller than the other approaches mentioned, leading to an increased participation of all involved surfaces with their oxides and contaminations and consequently also to higher losses. Comparing the VG-microstrip resonators to successful implementations of traveling-wave parametric amplifiers, however, shows that the achieved quality factors are indeed sufficient for this application and do in fact compare favorably (see Tab.\,\ref{tab:OtherWork}). As dielectric loss is one of the main suspects for noise exceeding the quantum limit in TWPAs \cite{esposito_perspective_2021}, we conclude that the fabrication method presented here is a promising candidate for their implementation.  

        \begin{table}
            \centering
            \caption{Loss comparison of our approach to successful implementations of TWPAs in parallel-plate geometry. }
            \label{tab:OtherWork}
            \small
            \vspace{0.3cm}
            \begin{tabular}{|c|c|c|}
            \hline
                 publication & dielectric & $\tan \delta$ \\\hline
                 Shan et al. \cite{shan_microwave_2015,shan_parametric_2016} &\SI{50}{\nano\metre}  \ch{SiO2} & \SI{5e-4}{} \\
                 Planat et al. \cite{planat_fabrication_2019,ranadive_kerr_2022,planat_photonic-crystal_2020,ranadive_traveling_2024}       & \SI{30}{\nano\metre} \ch{Al2O3} & \SI{6.5e-3}{} \\
                 Shu et al. \cite{shu_nonlinearity_2021,klimovich_demonstration_2023}  & \SI{200}{\nano\metre} \ch{aSi:H} & \SI{3.6e-5}{} \\\hline
                 this work &\SI{80}{\nano\metre}  \ch{vacuum} & \SI{3e-4}{} \\\hline
             \end{tabular}
             \vspace{-0.3cm}
        \end{table}

        To further investigate the origin of the losses of the resonators as well as to gain insight into the material properties of the aluminum and granular aluminum used as their center conductor, we explore the behavior of the resonators on Chip 1 with temperature. The fridge temperature is swept progressively in steps of \SI{50}{\milli\kelvin} up to \SI{700}{\milli\kelvin}. Beyond this temperature, none of the resonances can be fitted reliably anymore. 
        
\begin{figure}
            \centering
            \includegraphics[width=1.0\linewidth]{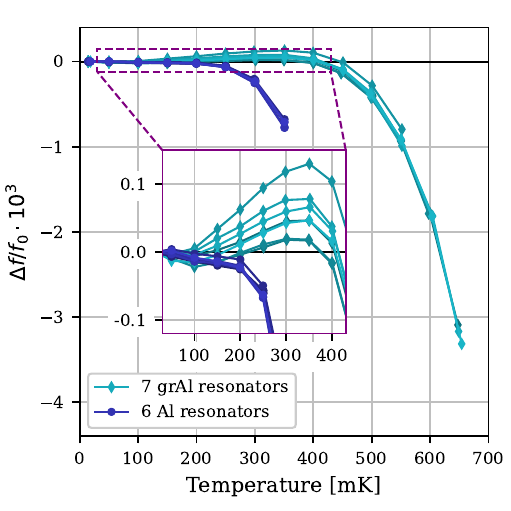} 
            \caption{Relative change in resonance frequency over temperature. Turquoise diamond markers resemble resonators made from granular aluminum, blue circles mark those made from pure aluminum. The inset shows a zoom in on the region between \SI{30}{\milli\kelvin} and \SI{430}{\milli\kelvin}. The lines connecting the markers are guides for the eye. The resonators in this experiment are the ones from Chip 1 (see Tab.\,\ref{tab:Resonators}).}
            \label{fig:Delta_f_over_T}
        \end{figure}

        With increasing temperature, we observe a decrease in internal quality factors as well as shifts in resonance frequency for all resonators. The corresponding data is shown in Fig.\,\ref{fig:Delta_f_over_T}. The temperature at which the resonance frequency starts to drop differs for aluminum and grAl resonators. The shift to lower frequencies for higher temperatures as well as the increase in frequency in the range \SIrange{100}{400}{\milli\kelvin} shown by the granular aluminum resonators can be explained by a combination of coupling to TLS in the dielectric and thermal quasi-particles according to the Mattis-Bardeen model \cite{pappas_two_2011,gao_physics_2008,tai_anomalous_2024,bruno_reducing_2015,grunhaupt_loss_2018}:
        
        \begin{widetext}
            \begin{align}
                \frac{f(T)-f_0}{f_0} = \frac{F\tan\delta}{\pi}\left(Re\left[\Psi\left(\frac{1}{2}+\frac{h f}{2i\pi k_\text{B}T}\right)\right]-\log \left(\frac{h f}{2\pi k_\text{B}T}\right)\right) -\frac{\alpha}{2}\sqrt{\frac{\pi\Delta_\text{S0}}{2k_\text{B}T}}\exp\left(\frac{-\Delta_\text{S0}}{k_\text{B}T}\right).
                \label{eq:fitfunc}
            \end{align}
        \end{widetext}

        Here $F\tan\delta$ is the loss tangent of the material connected to the TLS combined with its filling factor $F$, which, as we are only interested in the losses of our structures as a whole, can be set to 1. $\Psi$ denotes the complex digamma function. The kinetic inductance fraction $\alpha = L_\text{kin}/L_\text{tot}$ is the fraction of the kinetic inductance $L_\text{kin}$ of the resonator over the total inductance $L_\text{tot}$, which is the sum of the kinetic and geometric inductance. $\Delta_\text{S0}$ denotes the superconducting energy gap at zero temperature. It is worth noting that the effect of the TLS is more noticeable in the granular aluminum resonators, but also manifests in the aluminum resonators (see Supplementary Material \ref{sec:appendix_Temperature}). The more pronounced effect in the granular aluminum is due to two reasons. Firstly, the larger superconducting gap forces the frequency down at comparably higher temperature, revealing more of the TLS effect in the temperature range of \SIrange[]{150}{400}{\milli\kelvin}. As a second factor, it has to be mentioned that by design, the granular aluminum resonators reside in a lower frequency range than the aluminum resonators, namely $\SI{4}{\giga\hertz} < f_\text{grAl} < \SI{6.5}{\giga\hertz}$, compared to $\SI{7.5}{\giga\hertz} < f_\text{Al} < \SI{12}{\giga\hertz}$, which leads to the TLS part of eq.\,\ref{eq:fitfunc} having a larger effect at lower temperatures due to the factor $\frac{h f}{2\pi k_\text{B}T}$.
        
        Using eq.\,\ref{eq:fitfunc}, we can extract $\Delta_\text{S0}$ and $F\tan\delta$ by fitting it to the data presented in Fig.\,\ref{fig:Delta_f_over_T}. To obtain reliable fits, either the kinetic inductance fraction $\alpha$ or the superconducting gap $\Delta_\text{S0}$ has to be fixed. For the granular aluminum $\alpha = \SI{0.999(1)}{}$ is very close to unity because of the large kinetic inductance. Hence, also the uncertainty introduced by the exact value of the geometric inductance vanishes. The fitting results can be seen in Tab.\,\ref{tab:fitresults}. We obtain a loss tangent of $F\tan\delta = \SI{5.9(14)e-4}{}$ consistent with the obtained quality factors and a superconducting gap of $\Delta_\text{S0} = \SI{351(2)}{\micro\electronvolt}$. The values in brackets resemble the empirical standard deviation of the fit results of the individual data sets (cf. Supplementary Material \ref{sec:appendix_Temperature}). The value for the superconducting gap and the resulting critical temperature $T_\text{c} = 2\Delta_\text{S0}/(3.53 k_\text{B})=\SI{2.31(2)}{\kelvin}$ that we obtain for the granular aluminum lie within the range of values reported in the literature \cite{levy-bertrand_electrodynamics_2019,pracht_enhanced_2016,rotzinger_aluminium-oxide_2017}.
        \begin{table}
            \centering
            \caption{Parameters extracted by fitting eq.\,\ref{eq:fitfunc} to the data from Fig.\,\ref{fig:Delta_f_over_T}. For the granular aluminum resonators, the uncertainty given resembles the empirical standard deviation of the fits to the individual datasets. In contrast, for the aluminum, multiple different values were set for the superconducting gap $\Delta_\text{S0}$, and the resulting spread in the respective best fitting parameters is given.}
            \label{tab:fitresults}
            \small
            \vspace{0.3cm}
            \begin{tabular}{|c|c|c|}
                 \hline
                 & granular aluminum & aluminum \\
                 \hline
                 \begin{tabular}{@{}c@{}}number of \\ resonators \end{tabular}& 7&6 \\\hline
                  \begin{tabular}{@{}c@{}}fixed \\ parameters\end{tabular} &  $\alpha=\SI{0.999}{}$ & $\Delta_\text{S0}$ = \SI{210(20)}{\micro\electronvolt}\\
                  \hline
                 \begin{tabular}{@{}c@{}}fitted \\  parameters\end{tabular} & \begin{tabular}{@{}c@{}}$F\tan\delta$ = \SI{5.9(14)e-4}{} \\$\Delta_\text{S0}$=\SI{351(2)}{\micro\electronvolt} \end{tabular}& \begin{tabular}{@{}c@{}}$F\tan\delta$ = \SI{2.5(10)e-4}{}\\$\alpha = \SI{0.6(3)}{} $  \end{tabular}\\
                 \hline
             \end{tabular}
             \vspace{-0.3cm}
        \end{table}
        For the aluminum resonators, however, the kinetic inductance is expected to neither dominate nor vanish compared to the geometric inductance of the resonator. This is because of the geometric inductance $L_\text{geom}$ being suppressed by the close proximity of the suspended ground plane. Assuming the width $w$ of the center conductor being much larger than the vacuum-gap $h$ , we arrive at the approximate formula (see Supplementary Material\,\ref{sec:appendix_Inductance}):
        \begin{equation}
            \label{eq:Lgeom}
            L_\text{geom} = \mu_\text{0} \frac{h}{w}.
        \end{equation}  
        Consequently, we fix the superconducting gap for the aluminum resonators and fit the data with the resulting function. From trying different fits, we can conclude that only a superconducting gap in the range $\Delta_\text{S0} =\SIrange{190}{230}{\micro\electronvolt}$ agrees with the observed data (see Supplementary Material \ref{sec:appendix_Temperature}). A value of \SI{210(20)}{\micro\electronvolt} is consistent with the literature \cite{court_energy_2007} and also leads to a kinetic inductance fraction $\alpha = \SI{0.6(3)}{} $ consistent with our expectations from finite element electromagnetic simulations. The fit result for the TLS contribution $F \tan \delta = \SI{2.5(10)e-4}{}$ is consistent with TLS being the limiting factor of the resonators' quality factors.
        
        To conclude, we have developed a fabrication process for vacuum-gap microstrips with a narrow gap of \SI{80}{\nano\meter}, and used it to build distributed aluminum and granular aluminum $\lambda/4$ resonators. The fabrication process enables air-bridges, compact lumped-element capacitors, and high-capacitance transmission lines. It is worth pointing out that it is similar to the process developed in \cite{bruckmoser_niobium_2025}, with the distinction that in our work the achieved gap size is smaller by approximately a factor of \SI{30}{} and thereby the capacitance is higher. Of course, this comes at the cost of increased losses due to surface participation and resist residue on these surfaces. We have shown that two-level systems, likely situated in these thin surface layers, dominate the losses of our structures. The achieved quality factors compare favorably to parallel plate designs using a dielectric to separate the ground plane from the base layer, making our process promising for the fabrication of quantum-limited TWPAs. Only using low temperature steps and no aggressive chemicals, it is, in principle, compatible with almost all materials, notably including Josephson junctions as well as high-kinetic inductance materials such as \ch{NbTiN} and grAl. Therefore, the process is also compatible with superconducting qubits and low-speed-of-light transmission lines, possibly facilitating multiple qubit entanglement in a compact system. Finally, it makes it possible to realize compact on-chip capacitors and strong capacitive couplers, which are useful for the implementation of microwave detectors.

\section*{Supplementary Material}

See the supplementary material for details on the grAl sputtering (A), a list of paramters for the measured resonators (B), more details on the temperature dependence (C), details on calculating the inductance of the microstrip in the vacuum gap (D) and on the experimental setup (E) as well as an EDX analysis of the resonators. 

\begin{acknowledgments}
 This project has received funding from the Horizon Europe 2021-2027 project TruePA of the European Union (grant agreement number 101080152). RA was funded in part by the Austrian Science Fund (FWF) DOI 10.55776/F71. For the purpose of open access, the authors have applied a CC BY public copyright license to any Author Accepted Manuscript version arising from this submission.
\end{acknowledgments}

\bibliographystyle{apsrev4-1.bst}
\bibliography{biblio.bib}

\clearpage

\appendix

\begin{widetext}
\counterwithin{figure}{section}
\counterwithin{table}{section}
\textbf{Supplementary Material:Fabrication and characterization of vacuum-gap microstrip resonators}

\section{Granular aluminum sputtering}\label{sec:appendix_grAlSputtering}

    The granular aluminum used as the central conductor for the resonators in this work was deposited by DC sputtering of an aluminum target in an \ch{Ar} + \ch{O2} atmosphere. The gas flow into the chamber is controlled by two mass flow controllers, one dispensing pure \ch{Ar} gas, while the other one controls the flow of a \SI{95}{\percent} \ch{Ar} and \SI{5}{\percent} \ch{O2} gas mixture. This setup enables finer control and more stable application of the oxygen compared to the use of pure oxygen. Fine tuning of oxygen abundance, sputtering power, as well as annealing temperature after the process, facilitates control over the normal state resistance and thereby the kinetic inductance of the deposited film. Also, we found that flooding the chamber with oxygen before the sputter deposition leads to enhanced reproducibility of the process.

\section{Resonator measurements}\label{sec:appendix_QualityFactors}

    The data used for the assessment of the quality factors of the VG-microstrips presented in Fig.\,\ref{fig:Q_over_n} stems from resonators distributed over three transmission lines in two cool-downs. All data used to produce the plots can be seen in Fig.\,\ref{fig:Q_over_n_all_lines}. Of the 12 granular aluminum resonators that were cooled down, 11 could be observed. For the aluminum resonators, this number is 10 out of 15. All resonators used for this work are listed in Tab.\,\ref{tab:Resonators} with their resonance frequencies, coupling, and internal quality factors. We observe some deviations of the resonance frequencies from their designed values see Tab.\,\ref{tab:Resonators}. We attribute this mainly to deformations of the VG microstrips. A reduction or increase in the distance between the ground plane and the center conductor changes the capacitance as well as the geometric inductance (see appendix \ref{sec:appendix_Inductance}) and will thereby lead to a change in frequency. This could also explain why the granular aluminum resonators are shifted further from their design values than the aluminum resonators. For the aluminum resonators, the decrease in geometric inductance will partly compensate for the increase in capacitance that comes with a reduced distance between the center conductor and the ground plane (see Supplementary Material\,\ref{sec:appendix_Inductance}). Also, this is likely to explain the resonators that could not be observed, as a ground plane that is too strongly deformed might touch the center conductor of a resonator and create a short.

    \newcolumntype{t}{>{\columncolor{white!70!teal}}c}

    \begin{table}[h]
        \centering
        \caption{Measured resonance frequency $f_\text{res}$, coupling quality factor $Q_\text{c}$ and low-power internal quality factor $Q_\text{int}$ for all observed resonators. The respective design values $f_\text{res,des}$ and $Q_\text{c,des}$ are also presented.}
        \label{tab:Resonators}
        \small
        \vspace{0.3cm}
        \begin{tabular}{|c|c|c|c|c|c|c|c|}
        \hline 
             sample         & \begin{tabular}{@{}c@{}}resonator \\ number \end{tabular} & material & $f_\text{res} [\SI{}{\giga\hertz}]$   &  $f_\text{res,des} [\SI{}{\giga\hertz}]$  & $Q_\text{c}$   & $Q_\text{c,des}$   & $Q_\text{int}$      \\\hline
             Chip 1, line 1 & 1 & grAl & \SI{4.89}{ }& \SI{6.5}{} & \SI{2.0e3}{} & \SI{8.2e2}{} & \SI{3.9e3}{} \\
             Chip 1, line 1 & 2 & grAl & \SI{5.66}{ }& \SI{7.0}{} & \SI{1.3e3}{} & \SI{7.1e2}{} & \SI{3.1e3}{} \\
             Chip 1, line 1 & 3 & grAl & \SI{6.04}{ }& \SI{7.5}{} & \SI{1.6e3}{} & \SI{6.2e2}{} & \SI{3.8e3}{} \\

             Chip 1, line 1 & 4 & Al & \SI{7.55}{ }& \SI{8.0}{} & \SI{5.9e3}{} & \SI{9.6e3}{} &\SI{2.7e3}{} \\
             Chip 1, line 1 & 5 & Al & \SI{8.61}{ }& \SI{8.5}{} & \SI{3.7e3}{} & \SI{8.5e3}{} & \SI{2.9e3}{} \\
             Chip 1, line 1 & 6 & Al & \SI{9.33}{ }& \SI{9.5}{} & \SI{9.3e3}{} & \SI{6.8e3}{} &\SI{3.6e3}{} \\\hline
             
             Chip 1, line 2 & 7 & grAl & \SI{4.18}{ }& \SI{6.0}{} &\SI{3.1e3}{} & \SI{9.6e2}{} & \SI{2.3e3}{} \\
             Chip 1, line 2 & 8 & grAl & \SI{4.25}{ }& \SI{6.5}{} &\SI{4.0e3}{} & \SI{8.2e2}{} & \SI{2.9e3}{} \\
             Chip 1, line 2 & 9 & grAl & \SI{4.96}{ }& \SI{7.0}{} &\SI{2.7e3}{} & \SI{7.1e2}{}& \SI{2.9e3}{} \\
             Chip 1, line 2 & 10& grAl & \SI{5.34}{ }& \SI{7.5}{} &\SI{3.1e3}{} & \SI{6.2e2}{}& \SI{3.0e3}{} \\

             Chip 1, line 2 & 11 & Al & \SI{8.27}{ }& \SI{8.5}{}  &\SI{8.0e3}{} & \SI{8.5e3}{} &\SI{3.7e3}{} \\
             Chip 1, line 2 & 12 & Al & \SI{8.93}{ }& \SI{9.0}{}  &\SI{2.1e3}{} & \SI{7.5e3}{} & \SI{3.5e3}{} \\
             Chip 1, line 2 & 13 & Al & \SI{9.85}{ }& \SI{10.0}{} &\SI{8.5e3}{} & \SI{6.1e3}{} & \SI{4.4e3}{} \\\hline
             
             Chip 2, line 1 & 14 & grAl & \SI{3.92}{ }& \SI{6.0}{} & \SI{1.5e3}{} & \SI{9.6e2}{} & \SI{2.1e3}{} \\
             Chip 2, line 1 & 15 & grAl & \SI{4.30}{ }& \SI{6.5}{} & \SI{1.3e3}{} & \SI{8.2e2}{} & \SI{2.2e3}{} \\
             Chip 2, line 1 & 16 & grAl & \SI{4.75}{ }& \SI{7.0}{} & \SI{1.1e3}{} & \SI{7.1e2}{} & \SI{2.4e3}{} \\
             Chip 2, line 1 & 17 & grAl & \SI{5.04}{ }& \SI{7.5}{} & \SI{1.2e3}{} & \SI{6.2e2}{} & \SI{2.4e3}{} \\

             Chip 2, line 1 & 18 & Al & \SI{7.18}{ }  & \SI{8.0}{}  & \SI{15.9e3}{} & \SI{9.6e3}{} & \SI{1.9e3}{} \\
             Chip 2, line 1 & 19 & Al & \SI{8.59}{ }  & \SI{8.5}{}  & \SI{12.6e3}{} & \SI{8.5e3}{} & \SI{2.7e3}{} \\
             Chip 2, line 1 & 20 & Al & \SI{8.75}{ }  & \SI{9.0}{}  & \SI{7.2e3}{}  & \SI{7.5e3}{} & \SI{2.2e3}{} \\
             Chip 2, line 1 & 21 & Al & \SI{11.59}{ } & \SI{10.0}{} & \SI{140}{}    & \SI{6.1e3}{} & \SI{2.1e3}{} \\\hline
         \end{tabular}
         \vspace{-0.3cm}
    \end{table}

    \begin{figure*}[h]
        \centering
        \includegraphics[width=\textwidth]{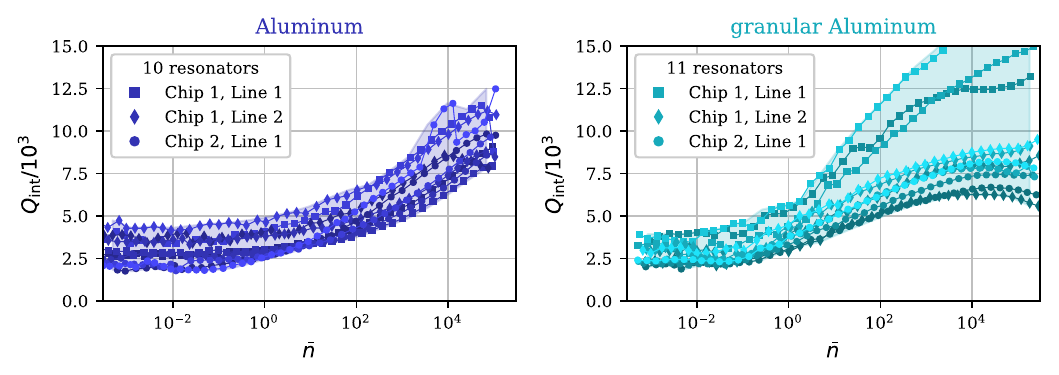} 
        \caption{Internal quality factors of aluminum \textbf{(left)} and granular aluminum \textbf{(right)} vacuum-gap microstrip resonators over average photon number $\Bar{n}$. The quality factors were obtained by circle-fitting~\cite{probst_efficient_2015} the resonator responses in the VNA transmission data of the feedline.}
        \label{fig:Q_over_n_all_lines}
    \end{figure*}
    
    The parallel-plate vacuum-gap couplers, coupling the resonators to the transmission line, are identical for all resonators. They span the transmission line as an air bridge with a width of \SI{5}{\micro\metre}. Because the coupling quality factor depends not only on the capacitance, but also on the resonance frequency and impedance of the resonator, this leads to different coupling quality factors for different resonators. From the design, we expect the aluminum resonators to couple with $\SI{5e3}{} < Q_\text{c} < \SI{10e3}{}$ and the granular aluminum resonators to couple with $\SI{0.6e3}{} < Q_\text{c} < \SI{1e3}{}$ because of their larger impedance. The circle fits yield $\SI{2e3}{} < Q_\text{c} < \SI{16e3}{}$ for the aluminum resonators with one very strongly coupled outlier at $Q_\text{c}=\SI{150}{}$ and $\SI{1e3}{} < Q_\text{c} < \SI{4e3}{}$ for the granular aluminum resonators. We attribute the deviations to standing waves in the CPW transmission line caused by impedance mismatch at the wire bonds connecting the chip to the PCB. Additionally, the already discussed deformation of the vacuum-gap structures can lead to different couplings. The one very strongly coupled aluminum resonator suggests an almost complete collapse of the corresponding coupler.

    To generate the plots in Fig.\,\ref{fig:Q_over_n} in the main text, we combined all the data into a compact depiction using the average as well as the minimum and maximum. The data is organized in equally spaced bins, the number of which (21 for the grAl and 20 for the aluminum resonators) is chosen to avoid artifacts of this method, such as fluctuations in the average curve due to sampling.

\section{Temperature dependence}\label{sec:appendix_Temperature}

In order to investigate the temperature dependence of the resonators, the base temperature of the dilution refrigerator is swept up to \SI{1}{\kelvin} in the cool-down containing Chip 1. The data of all resonators used to produce Fig.\,\ref{fig:Delta_f_over_T} is shown in Fig.\,\ref{fig:F_over_T_all_fits} together with the fits mentioned in the main text. The curves are shown with artificial offsets to improve visibility. 

    \begin{figure*}[h]
        \centering
        \includegraphics[width=\textwidth]{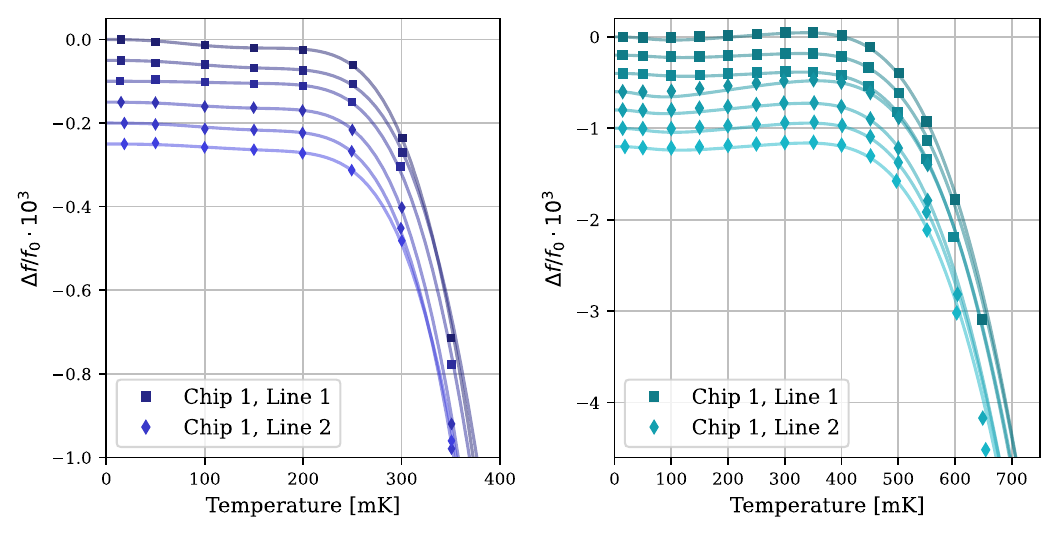} 
        \caption{Relative change in resonance frequency over temperature with best fits to eq.\,\ref{eq:fitfunc}. The data and fits of different resonators are artificially offset for better visibility. \textbf{(left:)} To fit the aluminum resonators, the superconducting gap was fixed. The fits shown here are for $\Delta_\text{S0} =  \SI{210}{\micro\electronvolt}$, but an uncertainty of $\pm\SI{20}{\micro\electronvolt}$ is estimated. \textbf{(right:)} For the granular Aluminum resonator, the kinetic inductance fraction is fixed to $\alpha=0.999$.}
        \label{fig:F_over_T_all_fits}
    \end{figure*}

    As the kinetic inductance fraction $\alpha$ is not well known for the aluminum film, the superconducting gap $\Delta_\text{S0}$ has to be fixed for the fit. Trying different plausible values for $\Delta_\text{S0}$, we arrive at the conclusion, that the behavior of the data in the region between \SIrange[]{100}{250}{\milli\kelvin} is best captured by $\Delta_\text{S0}=\SI{210}{\micro\electronvolt}$, but an uncertainty on that value of $\pm \SI{20}{\micro\electronvolt}$ is estimated. This uncertainty consequently leads to a spread in the extracted fit parameters. As this spread is larger than the empirical standard deviation of the collection of data sets, it is therefore given as the error on the fit results. 
    
    To further investigate the material parameters of the aluminum used for the center conductor of the resonators, the critical temperature of the specific thin-film as well as more dense data especially in the region \SIrange[]{100}{250}{\milli\kelvin} would be required, which was not the focus of this study. Also, resonators at a lower resonant frequency would show a more pronounced TLS effect at lower temperatures due to the nature of the TLS part of eq.\,\ref{eq:fitfunc}. The term $\frac{h f}{2i\pi k_\text{B}T}$ in the complex digamma function $\Psi$ as well as in the logarithm yields larger values for the term, shifting the relative frequency by a larger amount. This, together with the lower superconducting gap of aluminum as compared to granular aluminum, explains the qualitative difference that can be observed in the behavior of the two materials with temperature.

    In contrast to the aluminum resonators, the resonators made from grAl are heavily dominated by kinetic inductance due to the (\SI{100}{\pico\henry\per\sq}) kinetic inductance of the film and the suppressed geometric inductance stemming from the microstrip geometry (see \ref{sec:appendix_Inductance}). This allows us to fix it to \SI{99.9(1)}{\percent} with much larger accuracy compared to the aluminum resonators. Consequently, the empirical standard deviation between fitting parameters of different resonators is given as the error on the respective values.  

\section{Inductance of a microstrip}\label{sec:appendix_Inductance}

 To estimate the geometric inductance $L_\text{geom}$ of the VG-microstrips, we consider two parallel conducting plates with length $l$ separated by a distance $h$. The width $w$ of the bottom plate is less than that of the top ground plane. In the following, we will assume $l\gg w \gg h$. The two plates carry two currents $I$, equal and opposite in direction along the z-axis, for which we assume a homogeneous distribution along the plate in the region where the two plates overlap (see Fig.\,\ref{fig:inductance_sketch}). The resulting magnetic field is constant in the space between the plates. Applying Ampère's law to loops around the two plates individually, we obtain:

    \begin{figure}[h]
        \centering
        \includegraphics[width=0.5\linewidth]{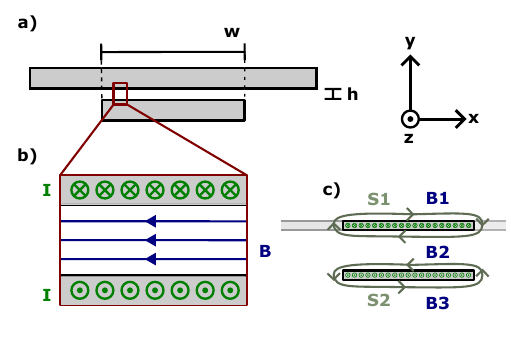}
        \caption{Sketch illustrating the calculation of the geometric inductance of a conductor in microstrip geometry.}
        \label{fig:inductance_sketch}
    \end{figure}

    \begin{align}
        \oint_{\text{S1}} \vec{B} d \vec{s} = w (-B_\text{2} + B_\text{1}) = \mu_\text{0}I  \label{eq:IntS1}\\
        \oint_{\text{S2}} \vec{B} d \vec{s} = w ( B_\text{3} - B_\text{2}) = \mu_\text{0}I \label{eq:IntS2}
    \end{align}
    Note that the direction of integration around the plates has been chosen in such a way as to match the direction of the current in the respective conductor. Due to the distance between the plates being much smaller than their width, the homogeneous fields created by the two current sheets cancel outside the arrangement, i.e., $B_\text{1}=B_\text{3}=0$. Therefore from equations \ref{eq:IntS1} and \ref{eq:IntS2} we obtain:
    \begin{equation}
        B_\text{2} = - \frac{\mu_\text{0}I}{w},
    \end{equation}
    meaning a $\vec{B}$-field in negative x-direction and with absolute value $|\vec{B}| = \mu_\text{0}I/w$. To find the inductance $L = \Phi_\text{m} / I$, we first find the magnetic flux $\Phi_\text{m}$ through a one unit length long cross section of the center space between the two plates
    \begin{equation}
        \Phi_\text{m} = \int_\text{0}^\text{1} \int_\text{0}^h \frac{\mu_\text{0}I}{w} dy dz = \frac{\mu_\text{0} h I}{w}.
    \end{equation}
    This directly yields the inductance per unit length:
    \begin{equation}
        L_\text{geom} = \mu_\text{0} \frac{h}{w}.
    \end{equation}
    The geometric factor $h/w$ makes clear how the geometry of the VG-microstrip suppresses the geometric inductance. If we plug in typical numbers for our resonators, we arrive at
    \begin{equation}
        L_\text{geom} \approx \SI{1,257e-6}{\newton\per\ampere\squared} \times \frac{\SI{80}{\nano\metre}}{\SI{4}{\micro\metre}} \approx \SI{25}{\nano\henry\per\metre}.
    \end{equation}
    This value agrees with the results of our finite element simulations within \SI{6}{\percent}. We attribute the remaining discrepancy mostly to the assumptions taken in this analytical approach, namely, possible fringe effects at the edges of the smaller one of the conducting sheets.

\section{Experimental setup}\label{sec:appendix_Setup}
    \begin{figure}[h]
        \centering
        \includegraphics[width=.8\linewidth]{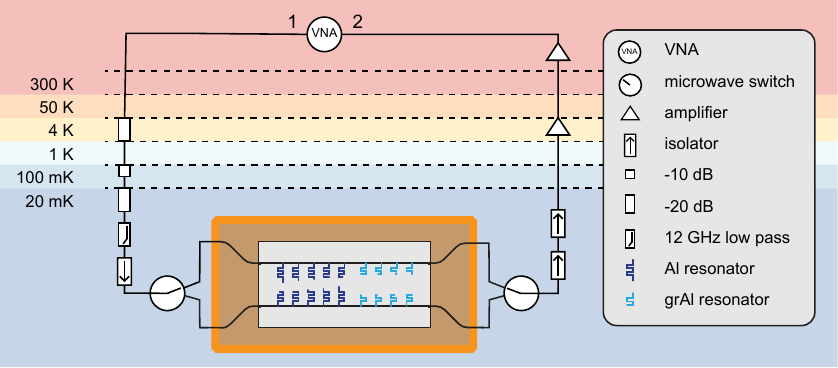}
        \caption{Sketch illustrating the measurement setup in the dilution refrigerator. Two microwave switches at the \SI{20}{\milli\kelvin} stage of the cryostat were used to switch between transmission lines.}
        \label{fig:setup}
    \end{figure} 

    The measurements presented in this work were performed in a dilution refrigerator setup depicted in Fig.\,\ref{fig:setup}. We placed the chip in the center of a printed circuit board (PCB) soldered to a copper sample holder and connected the on-chip bonding pads to the traces on the PCB using wire bonds. The copper sample holder was then mounted on the base plate of the refrigerator and connected to the microwave cabeling via SMA connectors. At room temperature, we use a Keysight VNA to measure the transmission through the system, with the signal going in via an attenuated input line, passing through the sample, and leaving the cryostat through a low-noise cryogenic amplifier as well as an amplifier at room temperature. The VNA power was swept from \SI{-100}{dBm} to \SI{0}{dBm}. It is worth noting that due to the nonlinearity of the resonators, the resonances at high powers could not be fitted reliably and are therefore not represented in this work. To determine the absolute power at the sample, we estimate the attenuation of the input line, including all components and cables in and outside the fridge to \SI{63}{dB}. Two microwave switches, visible in Fig.\,\ref{fig:setup} on both sides of the sample box, were used to switch between feedlines.

\section{EDX Analysis}\label{sec:appendix_EDX}
    We performed an EDX analysis to determine the degree to which the resist is removed from the VG-microstrips. Three samples were processed up to and including the etching of the ground plane (step c in Fig.\,\ref{fig:fabsketch}) before they were treated separately. Sample A was not exposed to organic solvents at all, sample B was left in NMP for \SI{5}{\hour}, and sample C was left in NMP for \SI{48}{\hour}. The NMP baths for samples B and C were followed by a solvent-removal sequence consisting of a bath in acetone, a bath in ethanol, and critical point drying in \ch{CO2}. After this treatment (summarized in Tab.\,\ref{tab:EDX_samples}), and for all three samples, the bulk of the ground plane was removed by \ch{BCl3} followed by \ch{CF4} plasma etching, leaving behind only the thin innermost aluminum layer. This is essential for resolving the EDX signal of the sacrificial resist within the VG-microstrips. We use the carbon within the resist as a marker for residue and surface contamination in the samples.

    \begin{figure*}[h]
        \centering
        \includegraphics[width=\textwidth]{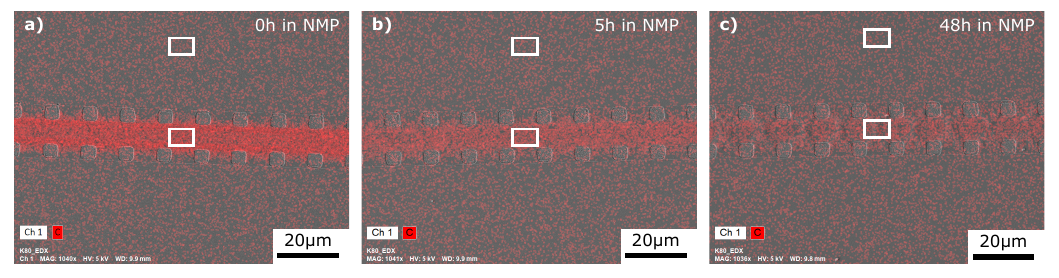} 
        \caption{EDX data for carbon, overlaid on SEM images of vacuum-gap microstrips. Samples A, B, and C from Tab.\,\ref{tab:EDX_samples} are shown in \textbf{a}, \textbf{b}, and \textbf{c}, respectively. The white rectangles mark the areas in which the raw EDX carbon count rates presented in Tab.\,\ref{tab:EDX_samples} were recorded.}
        \label{fig:EDX}
    \end{figure*}
    
    Fig.\,\ref{fig:EDX} shows the counts attributed to carbon on sections of VG microstrip overlayed on SEM images of samples A, B, and C. In addition to the presented maps, for each sample the count rates were taken in two zoomed-in areas on the VG microstrip and on the background respectively. White rectangles in Fig.\,\ref{fig:EDX} mark the corresponding areas; the resulting count rates in counts per second per \SI{}{\electronvolt} are presented in Tab.\,\ref{tab:EDX_samples}. The carbon count for the first sample, which did not experience any solvent cleaning, is the highest. The area covered by the VG microstrip (strip between the two rows of holes) is clearly distinguishable from the background on the sides. After \SI{5}{\hour} exposure to NMP, the contrast to the background is already reduced, but the outline of where the resist was is still present and there is still a change to be observed towards the sample that was in NMP for \SI{48}{\hour}. The count rates recorded in the zoomed-in areas support this visual impression, with the carbon count rate dropping by approximately half in the first \SI{5}{\hour} and then effectively halving again until the \SI{48}{\hour} mark. At this point, the carbon count rate is still raised from the background by about a factor of three.

     \begin{table}[h]
        \centering
        \caption{Samples used for the EDX analysis. All samples were subjected to the etching of the access holes before this treatment as well as to a removal of the ground plane in \ch{CF4} and \ch{BCl3} plasma afterwards. The EDX count rates were taken in zoomed-in regions directly on and next to the VG microstrips (marked with white rectangles in Fig.\,\ref{fig:EDX}). }
        \label{tab:EDX_samples}
        \small
        \vspace{0.3cm}
        \begin{tabular}{|c|c|c|c|c|c|c|}
        \hline 
             sample & NMP at \SI{70}{\degree}  & acetone        & ethanol        & CPD in \ch{CO2} & EDX VG microstrip  & EDX background \\\hline
             A      & --            & --             & --             & no & \SI{7.64}{cps\per\electronvolt} &\SI{0.57}{cps\per\electronvolt}\\
             B      &\SI{5}{\hour}  &\SI{24}{\hour}  & \SI{67}{\hour} & yes& \SI{3.52}{cps\per\electronvolt} &\SI{0.54}{cps\per\electronvolt}\\
             C      &\SI{48}{\hour} &\SI{24}{\hour} & \SI{24}{\hour}  & yes& \SI{1.85}{cps\per\electronvolt} &\SI{0.54}{cps\per\electronvolt}\\\hline
         \end{tabular}
         \vspace{-0.3cm}
    \end{table}

    We conclude from the EDX analysis that using \SI{48}{\hour} instead of \SI{5}{\hour} still decreases the amount of carbon residue present, and it can be assumed that longer exposure and potentially higher temperatures could lead to an even better result. Also, it is worth considering additional cleaning steps to remove the residue and improve the surface quality. Possible ways of achieving this could be to add a high-pressure oxygen plasma cleaning step after the critical point drying (CPD). To enhance the accessibility of the vacuum-gap for the plasma one can also consider a change in design, shifting from holes on the sides of the microstrips to slits across it. That way, the distance the plasma has to penetrate can be reduced without dramatically reducing the covered area.
 \end{widetext}   
\end{document}